\documentclass[aps,prb, preprintnumbers, twocolumn,showpacs,amsmath,amssymb,floatfix,eqsecnum,superscriptaddress]{revtex4}
\usepackage{graphicx}% Include figure files
\usepackage{dcolumn}% Align table columns on decimal point
\usepackage{bm}% bold math
\usepackage{psfrag}

\def\ket#1{\vert #1 \rangle}
\def\bra#1{\langle #1 \vert}
\def\me#1#2#3{\langle #1 \vert #2 \vert #3 \rangle}

\begin{document}

\title{Universal entanglement entropy in 2D conformal 
quantum critical points}

\author{Benjamin Hsu}
\affiliation{Department of Physics, University of Illinois at Urbana-Champaign, 
Illinois 61801-3080, USA}
\author{Michael Mulligan}
\affiliation{Department of Physics and SLAC, Stanford University, Stanford, California 94305, USA}
\author{Eduardo Fradkin}
\affiliation{Department of Physics, University of Illinois at Urbana-Champaign, 
Illinois 61801-3080, USA}
\author{Eun-Ah Kim}
\affiliation{Department of Physics, Cornell University, Ithaca, New York 14853, USA}
%\affiliation{Department of Physics, Stanford University, Stanford, California 94305}

\preprint{SU-ITP-08/30}
\preprint{SLAC-PUB-13470}

\date{\today}

\begin{abstract}
We study the scaling behavior of the entanglement entropy of two dimensional conformal 
quantum critical systems, {\it i.e.\/} systems with scale invariant wave functions. 
They include two-dimensional generalized quantum dimer models on bipartite lattices and  
quantum loop models, as well as the quantum Lifshitz model and related gauge theories.  
We show that, under quite general conditions, the entanglement entropy of a large and 
simply connected sub-system  of an infinite system with a smooth boundary has a universal finite contribution,
as well as scale-invariant terms %this term is not modular invariant, though
for special geometries. 
The universal finite contribution to the entanglement entropy is computable in 
terms of the properties of the conformal structure of the wave 
function of these quantum critical systems. 
The calculation of the universal term reduces to a problem in boundary conformal field theory.
\end{abstract}


\maketitle

\section{Introduction}
\label{sec:intro}

The non-local correlations of a quantum mechanical system are encoded in the behavior of the 
entanglement properties of its wave functions. A pure quantum state of a bipartite system $A\cup B$
 defines a mixed state in the observed region  $A$ obtained from tracing out the degrees of freedom in the unobserved region $B$. 
The non-local correlations connecting  regions $A$ and $B$ are encoded in the behavior of the
von Neumann entanglement entropy, $ S = - \textrm{Tr} \rho_{A} \ln \rho_{A}$, where $\rho_A$ 
is the reduced density matrix of region $A$. 
The entanglement entropy of a local quantum field theory, relativistic or not 
is known to exhibit an ``area law'' scaling of the form $S \sim \mu \ell^{D-1}$ in spatial dimensions $D>1$
where $\mu$ is a non-universal coefficient\cite{Srednicki1993,Bombelli1986}.
There has been growing interest in the scaling behavior 
of the  entanglement entropy at quantum critical points and in topological phases.
The entanglement entropy of  quantum critical systems in $D>1$ should contain universal  subleading terms, whose
structure for a general quantum critical system is not yet known.

The scaling behavior of the entanglement entropy has only been studied in detail 
in quantum critical systems in $D=1$ space dimension. Such systems are described by a
 $(1+1)$-dimensional conformal field theory (CFT).
In a $1+1$-dimensional CFT, 
 the entanglement entropy of a subsystem $A$ 
of linear size $\ell$  of an otherwise infinite system ({\it i.e.\/} of linear size $L \to \infty$) 
obeys a logarithmic scaling law,
\cite{Callan1994,Holzhey1994,Calabrese2004,Vidal2003,Latorre2004} 
$S\sim \frac{c}{3} \ln (\frac{\ell}{a})+\ldots$, where $c$ is the {\em central charge} of the CFT, and $a$ 
is the short distance cutoff. 
There has been a number of studies on topics related to this 1D  logarithmic scaling form. For instance,
a possible connection between this result and gravitational physics was suggested\cite{Ryu2006}. 
A similar logarithmic scaling behavior was found at infinite disorder fixed points of
1D random spin chains\cite{Refael2004,Refael2007}.
The quantum entanglement of quantum impurity systems has also been studied.
\cite{Kopp2007,Kopp2007a,Laflorencie2006,Sorensen2007,Affleck2008}
 
In this paper, we consider 
the universal scaling form of the 
entanglement entropy
 at 2D conformal quantum critical points (QCP) -- two-dimensional quantum critical systems 
 with scale-invariant many body wave functions. At a 2D conformal QCP,  
equal-time correlators of local operators coincide with 
 the correlation functions 
 of an appropriate 2D {\em classical} system
 at criticality (which is 
 described by an Euclidean 2D CFT)\cite{Ardonne2004}.  
 The entanglement entropy of 2D conformal QCPs was first considered in Ref.[\onlinecite{Fradkin2006}], where a 
 scaling form was found:
 $S=\mu \ell-\frac{c}{6} (\Delta \chi) \ln (\ell/a)+ \ldots$, where $c$ is the central charge of the 
 2D Euclidean CFT associated with the norm squared of the wave function and $\Delta \chi$ is the 
 change of the Euler characteristic $\chi$, $\Delta \chi=\chi_{A\cup B}-\chi_A-\chi_B$. 
 Notice for a region $A \subset B$ with a  smooth boundary, $\Delta\chi=0$ and hence the logarithmic term vanishes. 
 Hence, if region $A$ has a smooth boundary,
 there is no universal logarithmic term. In this case, we will show that instead there is a {\em finite},
${\cal O}(1)$, {\em universal term} $\gamma_{QCP}$ in the entanglement entropy at these quantum critical points, 
{\it i.e.\/}
 \begin{equation}
 S_{QCP}=\mu \ell + \gamma_{QCP}+\ldots.
 \label{eq:SQCP}
 \end{equation}
 Through explicit calculations and using general arguments based on CFT, we will show that $\gamma_{QCP}$
 has a topological meaning in the sense that  it is determined by the contributions of the winding modes of the underlying CFT.
  
 In a topological phase in 2D, the entanglement entropy scales as\cite{Kitaev2006a,Levin2006}
\begin{equation}
S_{\rm topo}=\alpha \ell -\gamma_{\rm topo} + O(\ell^{-1}),
\label{eq:Stopo}
\end{equation}
where $\alpha$ is a non-universal coefficient and 
$\gamma_{\rm topo}$, the {\em topological entanglement entropy}, is a topological invariant,    
the logarithm of the so-called total quantum dimension $\mathcal{D}
$ of the underlying topological field theory describing the topological phase.\cite{Kitaev2006a, Levin2006}  
Topological phases have non-trivial ground state degeneracies on 
surfaces of non-trivial topology. 
The topological entanglement entropy $\gamma_{\rm topo}$ also depends on the 
global topology of the manifold, and on surfaces with non-trivial topology, on the degenerate ground state on that
surface.\cite{Dong2008} 

Although superficially similar, the finite universal contributions to the entanglement entropy in topological phases 
and conformal quantum critical points, $\gamma_{\rm topo}$ and $\gamma_{QCP}$, have a different origin and structure. 
In the case of a topological phase, $\gamma_{\rm topo}$ is in general determined by the modular $S$-matrix of the 
topological field theory of the topological phase.\cite{Kitaev2006a,Levin2006,Dong2008} 
This modular $S$-matrix governs the transformation properties of the (degenerate) ground states of the topological phase 
on a torus under modular transformations, $\tau \to -1/\tau$, where $\tau$ is the modular parameter of the torus.\cite{Witten1989} 
However, we show below that for a general conformal quantum critical point, whose ground state wave function is given by the Gibbs weights of a Euclidean rational unitary CFT, the universal term $\gamma_{QCP}$ is determined by the modular $S$-matrix  associated with the norm squared of the wave function. Thus, the modular $S$-matrix of the topological phase and that of  the wave functions of 2D conformal quantum critical points have a conceptually different origin.  In particular, in all the cases we checked here, $\gamma_{QCP}$ and $\gamma_{\rm topo}$ contribute with opposite signs to their respective entanglement entropies, as implied by the conventions we used in Eq.\eqref{eq:SQCP} and Eq.\eqref{eq:Stopo}.

We will show that, when the logarithmic terms 
in the
entanglement entropy cancel, the
finite terms $\gamma_{QCP}$ are universal and are determined not only by the central charge but also by the restrictions on the states
imposed by the compactification conditions. 
Furthermore,  the form of the result for the entanglement entropy of Eq.\eqref{eq:SFM}  implies a 
connection with boundary CFT, as developed by Cardy.\cite{Cardy1986a,Cardy1986b} Thus,
 in addition of it  being determined by the central charge $c$, it must also depend on
the operator content of the CFT. For the same reason, the structure of Eq.\eqref{eq:SFM} 
also suggests a direct connection between this problem and the Affleck-Ludwig boundary entropy of 1D quantum CFTs.
\cite{Affleck1991} 

The paper is organized as follows. 
In Section \ref{sec:qdm} we apply this approach first to the simpler 
case of the quantum Lifshitz model (and the related quantum dimer models, QDMs) on  planar, cylindrical and toroidal geometries. 
These results apply to the QCPs of (generalized) quantum dimer model on bipartite lattices \cite{Rokhsar1988,Moessner2001,Fendley2002,Fradkin2004,Papanikolaou2007b,Castelnovo2005} and in  
   quantum eight-vertex models\cite{Ardonne2004}. 
Through explicit calculations for various geometries, we show that  
that, when the logarithmic terms 
in the entanglement entropy cancel, and that the subleading 
finite terms $\gamma_{QCP}$ are universal, determined not only by the central charge but also by the restrictions 
imposed by the compactification conditions. 
 In Section \ref{sec:generalizedqcps} we 
generalize this result to all 2D conformal QCPs whose scale-invariant wave functions have norms that are the partition functions of  2D Euclidean Rational CFTs (RCFT), CFTs
with a finite number of primary fields\cite{Ginsparg:1988nr,yellow}. 
More specifically, we show that the finite term in the
entanglement entropy of the 2D wave function is determined by the change of the Affleck-Ludwig boundary 
entropy of the 1D CFT -- a quantity determined by the modular $S$-matrix of the associated CFT and by the coefficients 
in the fusion
rules.
We also discuss specific examples of this class including 
2D quantum  loop models \cite{Freedman2004b} which, with the naive inner product, 
are known  to be quantum critical. \cite{Troyer2008,Fendley2008}
 We also briefly discess the quantum net models.\cite{Freedman2004b,Levin2005,Troyer2008,Fendley2008}
 In Section \ref{sec:conclusions} we conclude with a summary and a discussion on open questions. In particular, we comment on the implications of our results to the nature of related topological phases.

\section{Quantum Lifshitz model universality class}
\label{sec:qdm}

The quantum Lifshitz model\cite{Ardonne2004} (QLM) in two space dimensions
is defined by the following Hamiltonian with an arbitrary paramter $k$:
\begin{equation}
	H = \int d^{2} x \left[ \frac{\Pi^2}{2} +\frac{1}{2} \left(\frac{k}{4\pi}\right)^2 (\nabla^{2} \phi)^{2} \right],
	\label{eq:qlm}
\end{equation}
where $\phi$ is a scalar field   $\Pi = \dot{\phi}$ is its canonical momentum conjugate to $\phi$. The QLM Hamiltonian Eq.\eqref{eq:qlm} defines a class of  
 QCP's  with dynamic critical exponent $z=2$, and  a continuous parameter $k$.

This remarkable property of the model is evident in the exactly known wave function for the ground state $|GS\rangle$ which is a superposition of all field configurations $\phi(x,y)$
with the configuration dependent weight\cite{Ardonne2004}: 
\begin{equation}
	\Psi_{GS}[\phi] = \langle[\phi]|GS\rangle=\frac{1}{\sqrt{Z} }   
		e^{\displaystyle{-S[\phi]/2}},
		\label{eq:Psi0qlm}
\end{equation}
with
\begin{equation}
S[\phi]=\int d^2x \; \frac{k}{4\pi} \left({\vec \nabla} \phi(x)\right)^2
\label{eq:Sofphi}
\end{equation}
and the norm squared of the state
\begin{equation}
Z=||\Psi_{GS}||^2=\int D\phi \;\; e^{\displaystyle{-S[\phi]}}.
\label{eq:norm}
\end{equation}
Notice $Z$ is identical to 
the partition function for the Gaussian model, which defines free boson Euclidean CFT\cite{Nienhuis1987}, albeit with  the ``stiffness'' $k$. 
Hence Eq.\eqref{eq:qlm} defines an infinite class of 2D conformal QCP's all associated with free boson CFTs.

The QLM can be viewed low energy effective field theory capturing universal aspects of various microscopic lattice models
 with $\phi$ playing the role of coarse grained height field\cite{Moessner2002,Henley1997,Ardonne2004} with the ``stiffness'' $k$
 determined by  the appropriate ``microscopic'' coupling constants\cite{Ardonne2004,Papanikolaou2007b}.  
For such a mapping to work, the constraints of the lattice models should be build in through compactification of the boson field $\phi$ by demanding all physical operators to be invariant under the shift of $\phi\rightarrow \phi+2\pi r$ or equivalently all physical operators to take the form of vertex operators $e^{in\phi/r}$ for integer $n$.  
In subsection \ref{subsec:micro} we will discuss specific examples of this mapping corresponding to particular values of $k$ using the convention of fixing $r=1$.
The examples will include so-called Rokhsar-Kivelson point (RK) of the quantum dimer model\cite{Rokhsar1988} and its generalizations 
\cite{Castelnovo2005,Alet2005,Papanikolaou2007}
and  the {\em quantum} eight-vertex model\cite{Ardonne2004} 
special choices of the Baxter weight\cite{Baxter1982}.
Since $k$ can be varied in the QLM, this theory has an exactly {\em marginal} operator, resulting in continuously varying 
critical exponents (scaling dimensions) of the allowed (vertex) operators.\cite{Papanikolaou2007}

\subsection{Entanglement entropy and partition functions for 2D conformal QCPs}
\label{sec:FM-summary}

To investigate the universal finite  terms in the entanglement entropy at 
2D conformal QCPs, we will rely on the approach described in the work of Fradkin and Moore.\cite{Fradkin2006} 
They showed that $\textrm{tr} \rho_A^n$, where $\rho_A$ is
the (normalized) reduced density matrix of a
region $A$, with $A \subset B$ separated by the boundary $\Gamma$,
for the ground state $\Psi_0$ on $A \cup B$, 
is given by
\begin{equation}
\textrm{tr}\rho_A^n=  \frac{Z_n}{Z^n} =  \left(\frac{Z_A Z_B}{Z_{A \cup B}}\right)^{n-1}.
\label{eq:rhoA^n}
\end{equation} 
Here $Z_n$ is the partition function of $n$ copies of the equivalent 2D classical statistical mechanical 
system 
satisfying the constraint
that their degrees of freedom are identified on the boundary $\Gamma$, and
$Z^n$ is the partition function for $n$ decoupled systems. The partition functions on the  r.h.s of 
Eq.\eqref{eq:rhoA^n} are 
$Z_A=||\Psi_0^A||^2$ with support on region $A$ and $||\Psi_0^B||^2$ with support in region $B$, 
both satisfying 
generalized Dirichlet ({\it i.e.\/} fixed) boundary conditions on $\Gamma$ of $A$ and $B$, and 
$Z_{A \cup B}=||\Psi_0||^2$ is the norm squared for the full system.
The entanglement entropy $S$ is then obtained by an analytic continuation in $n$,  
\begin{eqnarray}
S&=&-\textrm{tr}
\left(\rho_A \ln \rho_A\right) \nonumber \\
&=& - \lim_{n \to 1} \frac{\partial}{\partial n} 
\textrm{tr} \rho_A^n \nonumber \\
&=& - \log \left(\frac{Z_A Z_B}{Z_{A \cup B}}\right)\nonumber\\
&&
\label{eq:SFM}
\end{eqnarray}
Hence, the computation of the entanglement entropy is reduced to the computation of a ratio of 
partition functions in a 2D classical
statistical mechanical problem, an Euclidean CFT in the case of a critical wave function, 
 each satisfying specific boundary conditions.

In order to construct  $\textrm{tr} \rho_A^n$,
  we need an expression for the matrix elements of the reduced density matrix 
  $ \me{\phi^A}{\rho_A}{{\phi^\prime}^A}$. 
Since the ground state wave function Eqs.\eqref{eq:Psi0qlm} and \eqref{eq:Sofphi}
 is a local function of the field $\phi(x)$, 
a general matrix element of the reduced density matrix is a trace of the density
matrix of the pure state $\Psi_{GS}[\phi]$ over the degrees of freedom of the ``unobserved'' region $B$, 
denoted by $\phi^B(x)$. Hence the matrix elements of $\rho_A$ take the form
\begin{eqnarray}
&&\me{\phi^{A}}{ \hat{\rho}_{A}}{ {\phi^\prime}^A    }
= \nonumber \\
&&\frac{1}{Z} \int [D\phi^{B} ] \,\, e^{\displaystyle{-\left(\frac{1}{2} S^{A}(\phi^{A}) + 
\frac{1}{2} S^{A}({\phi^\prime}^A ) 
+S^B(\phi^B)\right)}},
\nonumber \\
&&
\end{eqnarray}
where the degrees of freedom satisfy the {\em boundary condition} at the common boundary
 $\Gamma$:
\begin{equation}
BC_\Gamma:\quad \phi^B|_\Gamma=\phi^A|_\Gamma={{\phi^\prime}^A}|_\Gamma.
\label{eq:BCphiGamma}
\end{equation}
Proceeding with the computation of 
$\textrm{tr}\rho_A^n$, it is immediate to see that the matrix product  requires the condition  $\phi^A_i={\phi^\prime}^A_{i-1}$
 for $i=1,\cdots,n$, and ${{\phi^\prime}^A_n}=\phi^A_1$ from the trace condition. 
Hence, $\textrm{tr}{\rho_A^n}$ 
takes the form
\begin{eqnarray}
\textrm{tr} \rho_A^n&\equiv& \frac{Z_n}{Z^n}
\nonumber \\
&=&\frac{1}{Z^n} \int_{BC_\Gamma} \prod_i D \phi_i^A D\phi_i^B \; e^{\displaystyle_{-\sum_{i=1}^n
\left(S(\phi_i^A)+S(\phi_i^B)\right)}}
\nonumber \\
&&
\label{eq:trrhoAn1}
\end{eqnarray}
{\em subject to the boundary condition $BC_\Gamma$} of Eq.\eqref{eq:BCphiGamma}. 
Notice that the numerator, $Z_n$ is the partition
function on $n$ systems whose degrees of freedom are identified in $\Gamma$ but are 
otherwise independent. Also notice the absence of the factors of $1/2$ in the exponentials 
of Eq.\eqref{eq:trrhoAn1}.

The other important
consideration is that the compactification condition requires that two fields that differ by 
$2\pi r$ be equivalent. Hence, the
boundary condition of Eq.\eqref{eq:BCphiGamma} is defined {\em modulo $2\pi r$}. 
(Equivalently, the proper form of the
degrees of freedom is $e^{i\phi}$.) This means that one can alternatively define 
$Z_n$ as a partition function for $n$ 
systems which are decoupled {\em in the bulk} but have a boundary coupling of the 
form (in the limit $\lambda_\Gamma \to \infty$,
which enforces the boundary condition)
\begin{equation}
S_\Gamma=-\oint_\Gamma \lambda_\Gamma \sum_{i=1}^n
\cos(\phi_i-\phi_{i+1}).
\label{eq:SGamma}
\end{equation}
 Here the fields $\phi_i$ extend over the entire region $A \cup B$. 
Thus, this problem maps onto a boundary CFT for a system with $n$ ``replicas'' 
coupled only through the boundary condition on the closed contour $\Gamma$, the boundary 
between the $A$ and $B$ regions.

For the special case of the free scalar field, one can simplify this further by taking 
linear combinations of the replica fields. 
Then the condition that the scalar fields $\phi_{i}$ agree with each other on $\Gamma$ 
can be satisfied by forming $n-1$ relative 
coordinates $\varphi_i\equiv  \phi_{i}-\phi_{i+1}$ ($i=1,\ldots,n-1$) 
that vanish ({\em mod} $2\pi r$) on $\Gamma$, and one ``center of mass 
coordinate'' field $\phi\equiv \frac{1}{\sqrt{n} } \sum_{i=1}^n \phi_{i} $ 
that is unaffected by the boundary $\Gamma$ (reflecting the fact that nothing physical 
takes place at $\Gamma$).
Hence, the computation of $\textrm{tr} \rho_A^n$ reduces  to the product of two partition functions: 
\begin{enumerate}
\item
 The partition function for the ``center of mass'' field $\phi$; since $\phi$ does not see the 
 boundary $\Gamma$, this is just the partition function $Z_{A \cup B}$ for a single field in the 
 entire system. 
 \item
  The partition function for the $n-1$ fields $\varphi_i$ which are independent from each other 
  and vanish ({\em mod} $2\pi r$ on $\Gamma$. We denote this by $\left({Z^D_\Gamma}\right)^{n-1}$. 
  However,  the fields $\varphi_i$ on the $A$ and $B$ regions are effectively decoupled from each other. 
  Hence, this partition function further factorizes to $Z^D_\Gamma=Z_A^D Z_B^D$, where $Z_A^D$ and $Z_B^D$ 
  are the partition functions for a single field $\phi$ on $A$ and $B$ respectively, 
  satisfying in each case Dirichlet (fixed) boundary conditions ({\em mod} $2\pi r$) 
  at their common boundary $\Gamma$.
  \end{enumerate}
  Thus, we can write the trace $\textrm{tr} \rho_A^n$ as
\begin{equation}
\textrm{tr} \rho^{n}_{A}  = \frac{\left(Z^{D}_\Gamma\right)^{n-1} Z_{A \cup B}}{Z^{n}_{A \cup B}} = 
\left( \frac{Z^{D}_\Gamma }{Z_{F} }\right)^{n-1} = 
\left( \frac{ Z_{A}^D Z_{B}^D }{Z_{A \cup B} } \right)^{n-1}.
\label{eq:trrho^n}
\end{equation}
Here the denominator factor, $Z_{A \cup B}^n$ comes from 
the normalization factors, and represents the partition function over the entire system. 
The entanglement entropy is then\cite{Fradkin2006}
 \begin{equation}
 S = -\log Z_{A}^D - \log Z_{B}^D+ \log Z_{A \cup B}\equiv F_A^D+F_B^D-F_{A \cup B},
 \label{eq:SFM2}
 \end{equation}
 which, as indicated in the r.h.s of Eq. \eqref{eq:SFM2} reduces to the computation of the free energies 
 $F_A^D$, $F_B^D$ and $F_{A \cup B}$, for the equivalent 2D Euclidean CFT on regions $A$ and $B$, 
 each satisfying Dirichlet (fixed) boundary conditions on the common boundary $\Gamma$, and on the 
 full system, $A \cup B$, respectively.

The behavior of the free energy of a CFT as a function of the system size $\ell$ has been studied in detail. 
The divergent terms, as $\ell \to \infty$, have the form\cite{Kac1966, privman88, Cardy-Peschel1988}
\begin{equation}
F(\ell)=f_0 \ell^2+ \sigma \ell -\frac{c}{6} \chi \ln \left(\frac{\ell}{a}\right) + {\cal O}(1)
\label{eq:cardy-peschel}
\end{equation}
provided the boundary $\Gamma$ is smooth (and differentiable). Here, $f_0$ and $\sigma$ are two 
non-universal quantities, and $a$ is the short-distance cutoff; $c$ and $\chi$ are, respectively, 
the central charge of the CFT and the Euler characteristic of the manifold. 
It follows from this result  that the entanglement entropy for region $A$ takes the form\cite{Fradkin2006}
\begin{equation}
S=\alpha \ell - \frac{c}{6} (\Delta \chi) \ln \left(\frac{\ell}{a}\right)+{\cal O}(1).
\label{eq:FM06}
\end{equation}
provided the boundary $\Gamma$ is smooth. 
In all the geometries we discuss, the change in the Euler characteristic 
vanishes, $\Delta \chi=0$, and there is no logarithmic term. However we will show below that, if the logarithmic terms cancel, there exist a universal finite $O(1)$ term, 
as well as other universal 
dependences on the geometry (such as aspect ratios). 
We will now extract these universal finite  terms. 

\subsection{The Entanglement Entropy of the Quantum Lifshitz Universality Class}
\label{sec:entropy-qlm}

Here we calculate $\gamma_{QCP}$ at QCPs of the QLM universality class defined by Eq.\eqref{eq:qlm}
 for three different geometries: (i) a cylindrical geometry, (ii) a toroidal geometry,
 and (iii) a disk geometry. For the cylinder and disk we assume the 
 Dirichlet boundary conditions at the open ends.
 We use the known results on the free boson partition function\eqref{eq:norm}
 for different topologies and boundary conditions\cite{Polchinski86,Weisberger87,Ginsparg:1988nr,Fendley1994,Eggert1992,yellow},  
 which are necessary for the calculation of entanglement entropy.
It is useful to note that the action Eq.\eqref{eq:Sofphi}  for general value of the ``stiffness'' k turns into the standard form:
\begin{equation}
S[\varphi]= \frac{1}{8\pi} \int d^2x \; \left(\partial_\mu \varphi \right)^2,
\label{eq:rescaling}
\end{equation}
upon a rescaling of the field  $\sqrt{2k} \phi=\varphi$. If  $\phi$ is compactified with radius $r=1$,  
the rescaled field $\varphi$ has an effective compactification radius $R=\sqrt{2 kr^2}$.  
We find $\gamma_{QCP}$ to depend linearly on $\ln R$ in all cases we consider.

\subsubsection{The Cylinder}
\label{subsec:Cylinder}
%%%%%%%%%%%%%%%%%%%%%%%%%%%%%%
\begin{figure}[hb]
\begin{center}
\includegraphics[width=0.5\textwidth]{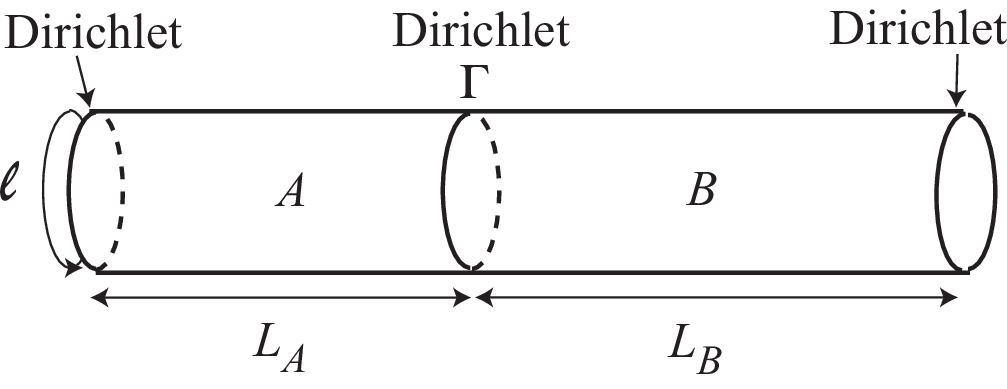}
\end{center}
\caption{Cylinder}
\label{fig:Cylinder}
\end{figure}

Let us begin by considering first a system on a long cylinder of linear size $L$ and circumference $\ell$ with
$L \gg \ell$. 
Region $A$ to be observed, is a cylinder of length $L_A$ and circumference 
$\ell$. The complement region, $B$, 
is a cylinder of length $L_B$ (see Fig.\ref{fig:Cylinder}), also with circumference $\ell$. We assume that the QLM wave function Eq.\eqref{eq:Psi0qlm}
 and hence the associated 2D partition function Eq.\eqref{eq:norm}
obey the Dirichlet boundary conditions at both ends of the cylinder, $A \cup B$. 

From Eq.\eqref{eq:SFM2}, the entanglement entropy $S_A=S_B\equiv S$ is given by
\begin{equation}
S= -\ln Z^{A}_{DD}(L_A,\ell)-\ln  Z^{B}_{DD}(L_B,\ell) +\ln Z^{A\cup B}_{DD}(L_A+L_B,\ell)
\label{eq:S-cyl}
\end{equation}
Here $Z_{DD}(L,\ell)$ is the partition function of Eq.\eqref{eq:norm} for  a boson with  compactification radius $R$ on cylinder of length $L$ and circumference $\ell$ with Dirichlet boundary conditions on both ends, which is well known:\cite{Fendley1994} 
\begin{equation}
Z_{DD}(L,\ell)=\mathcal{N}\; \frac{1}{R} \frac{\vartheta_3\left(\frac{2\tau}{R^2}\right)}{\eta(q^2)}
\label{eq:ZDD-cylinder}
\end{equation}
where $R=\sqrt{2r^2k}$ is the effective compactification radius (as before), and $\mathcal{N}$ is a non-universal regularization-dependent prefactor, responsible for the area and 
perimeter dependent terms in the free energy shown in Eq.\eqref{eq:cardy-peschel}. 
(There are no logarithmic terms for a cylinder or a torus as their Euler characteristic $\chi$ vanishes.)
In Eq.\eqref{eq:ZDD-cylinder} $\tau=i\frac{L}{\ell}$ is the modular parameter, encoding the geometry of the cylinder, 
and $q=e^{2\pi i \tau}$. The elliptic theta-function $\vartheta_3(\tau)$ and the Dedekind eta-function $\eta(q)$
are given by
\begin{equation}
\vartheta_3(\tau)=\sum_{n=-\infty}^\infty q^{\frac{n^2}{2}}, \quad
\eta(q)=q^{\frac{1}{24}} \prod_{n=1}^\infty (1-q^n).
\label{eq:theta3-eta}
\end{equation}
The important feature of Eq.\eqref{eq:ZDD-cylinder} is  the factor $1/R$, the contribution of the winding modes of the compactified boson on the cylinder 
with Dirichlet boundary conditions. 

Putting it all together, it is straightforward  to find an expression for the entanglement entropy using Eq.\eqref{eq:SFM}. 
In general, the entanglement entropy depends on the geometry ({\it e.g.\/} the aspect ratios $L/\ell$) of the cylinders, 
encoded in ratios of theta and eta functions. However, 
in the limit $L_A\gg \ell$, in which  the length of the cylinders are long compared to their circumference, 
the entanglement entropy given by Eq.\eqref{eq:S-cyl} and Eq.\eqref{eq:ZDD-cylinder} takes a simple form
\begin{equation}
S=\mu \ell +  \ln R,
\label{eq:entropy-cylinder}
\end{equation}
where $\mu$ is a non-universal constant that depending on the regularization-dependent pre-factor $\mathcal{N}$
of Eq.\eqref{eq:ZDD-cylinder}.  Hence, there is a $\mathcal{O}(1)$ universal contribution to the entanglement entropy 
 $\gamma_{QCP}= \ln R$ for the cylinderical geometry.  
The explicit dependence of $\gamma_{QCP}$ on the effective effective compactification radius $R=\sqrt{2kr^2}$ shows that it is determined by the winding modes of the compactified boson 
and thus it  is a universal quantity determined by the topology of the surface.
In particular we find that the universal piece of the entanglement entropy, $\gamma_{QCP}$, for a compactified boson is a 
 continuous function of the radius $R$, a consequence of the existence of an exactly marginal operator at this QCP. 
 We find the similar relations for all topologies we considered. 
We will come back to this point in section \ref{subsec:micro}, in the context of several microscopic models of interest.

\subsubsection{The Torus}
%%%%%%%%%%%%%%%%%%%%%%%%%%%%%%%%

\begin{figure}[hb]
\begin{center}
\includegraphics[width=0.5\textwidth]{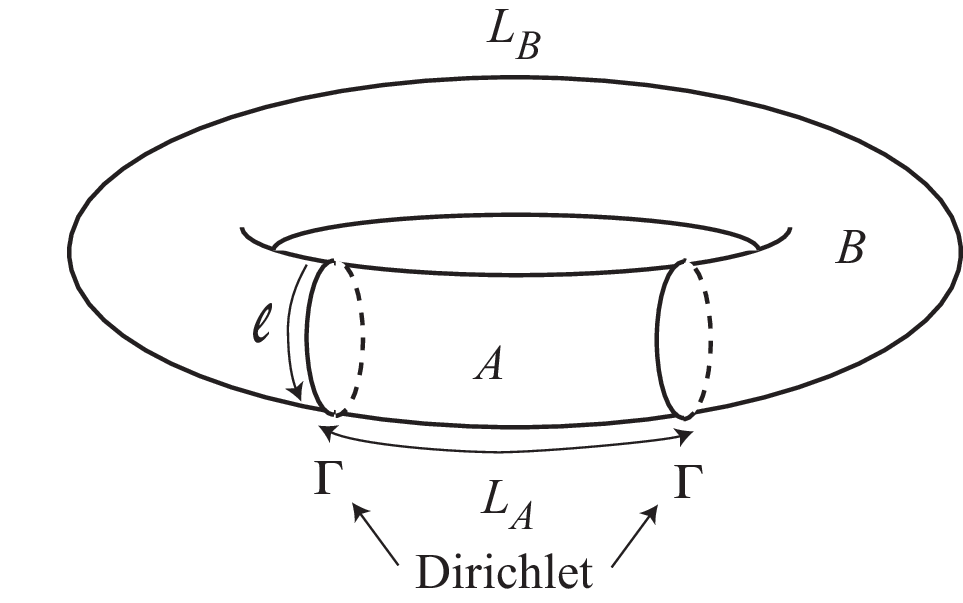}
\end{center}
\caption{Torus}
\label{fig:Torus}
\end{figure}

We now consider the case in which the full system $A\cup B$ is a torus for which the real part of the modulus $L/\ell \gg 1$, 
as shown in Fig.\ref{fig:Torus}. The two subsystems, $A$ and $B$ are now two cylinders, 
of length $L_A$ and $L_B$ respectively ($L=L_A+L_B$), 
both with the same circumference $\ell$. We will thus need the partition function 
on a torus and on two cylinders (with both ends of the cylinders obeying Dirichlet boundary conditions.) 
The trace $\textrm{tr} \rho_A^n$ now becomes
\begin{equation}
	\textrm{tr } \rho_{A}^{n} = 
	\left( \frac{Z^{A}_{DD}(L_A,\ell) Z^{B}_{DD}\left(L_B,\ell\right)}{Z^{A\cup B}_{\rm torus}(L,\ell) } \right)^{n-1}. 
\end{equation}
The partition functions for the two cylinders, $A$ and $B$ has the form of Eq. \eqref{eq:ZDD-cylinder}. 
The partition function
for the torus is\cite{yellow,Ginsparg:1988nr} 
\begin{equation}
Z_{\rm torus}(L,\ell) = \left(Z_{\rm cylinder}^{NN}\left(\frac{L}{2},\ell \right)\right)^2,
\label{eq:torus-cylinder}
\end{equation}
where $Z_{\rm cylinder}^{NN}(\frac{L}{2},\ell)$ is the partition function on a cylinder of length $\frac{L}{2}$ and
circumference $\ell$, with Neumann boundary conditions at both ends:
\begin{equation}
Z_{\rm cylinder}^{NN}\left(\frac{L}{2},\ell \right)=\mathcal{N}\; \sqrt{\frac{kr^2}{2}}\; \frac{\vartheta_3\left(\tau k r^2\right)}{\eta(q^2)},
\label{eq:}
\end{equation}
where $\tau=i\frac{L}{\ell}$ and $q=\exp(2\pi i \tau)$. 

In the limit $L_A \gg \ell \gg a$ and $L_B \gg \ell \gg a$, the entanglement entropy for the toroidal geometry is
\begin{equation}
S=\mu \ell +
2 \ln \left(\frac{R^2}{2}\right).
\label{eq:entropy-torus}
\end{equation}
Hence, for the toroidal geometry, the universal term is $\gamma_{QCP}=2 \ln \left(kr^2 \right)=2 \ln (R^2/2)$. 
In Eq.\eqref{eq:entropy-torus} $\mu$ is, once again, a non-universal factor which depends on both the short distance regularization and boundary
conditions (in fact, it is not equal to the constant we also called ``$\mu$'' in the entanglement entropy 
for the case of the
cylinder, Eq.\eqref{eq:entropy-cylinder}.)  As was the case for the cylindrical geometry, 
in the case of the torus
$\gamma_{QCP}$ is also determined by the contribution of the zero modes of the compactified boson to the partition functions.
Thus, here too, $\gamma_{QCP}$ depends on the 
effective boson radius $R=\sqrt{2kr^2}$. However, the different values of 
$\gamma_{QCP}$ in Eq.\eqref{eq:entropy-torus} and Eq.\eqref{eq:entropy-cylinder}
is due to the fact that on the torus all
three partition functions have contributions from the zero modes. 

\subsubsection{The Disk}
\label{subsec:Disk}
%%%%%%%%%%%%%%%%%%%%%%%%%%%%%%%%
\begin{figure}[hb]
\begin{center}
\includegraphics[width=0.4\textwidth]{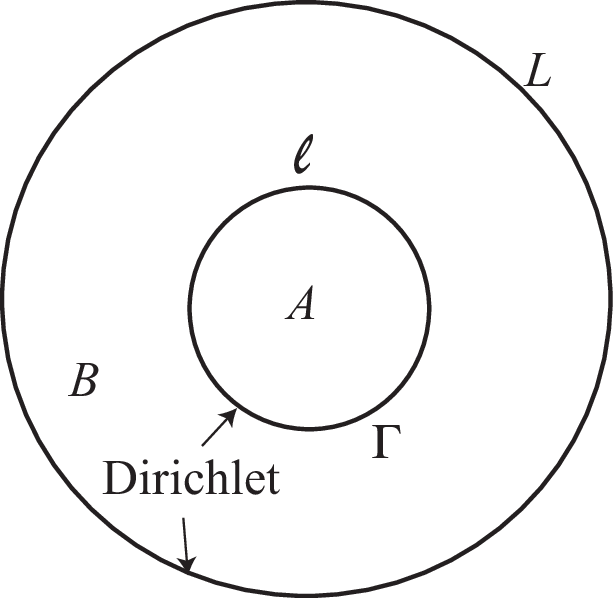}
\end{center}
\caption{Disk}
\label{fig:Disk}
\end{figure}
%%%%%%%%%%%%%%%%%%%%%%%%%%%%
Finally, we compute the entanglement entropy for the disk geometry, shown in Fig.\ref{fig:Disk}. 
The line of argument used
above applies here as well. This is the case discussed in Ref.[\onlinecite{Fradkin2006}], 
where it was found that the logarithmic
term in the entanglement entropy cancels exactly if the boundary $\Gamma$ is smooth. Here we compute the (subleading)
finite universal piece. 

To compute the entanglement entropy we need to compute three partition functions, on the two disks $A$ and 
$A \cup B$, 
and on the annulus $B$, all with Dirichlet boundary conditions. These partition functions were computed in the literature 
long
ago for an uncompactified boson.\cite{Polchinski86,Weisberger87} They can be obtained from the partition functions on 
cylinders, 
with Dirichlet-Dirichlet (for the annulus) and
Dirichlet-Neumann (for the disks) boundary conditions by a conformal mapping $w=\frac{\ell}{2\pi} \ln z$, from the $z$ 
complex
plane to the cylinder (labeled by $w$). The partition function for the annulus (region $B$) of inner circumference $\ell$ 
and
outer circumference $L$ (with Dirichlet boundary conditions) is
\begin{equation}
Z_{DD}^B(L,\ell)=\mathcal{N} \; \sqrt{\frac{\pi}{\ln \left(L/\ell\right)}}\;
\frac{1}{\sqrt{2kr^2}}\frac{\vartheta_3\left(\frac{\tau_B}{r^2k}\right)}{\eta(q_B^2)}.
\label{eq:Z-annulus}
\end{equation}
Except for the factor of $1/\sqrt{2kr^2}$, which is due to the zero modes of the compactified boson, this result agrees with
those of Ref.[\onlinecite{Weisberger87}]. In Eq.\eqref{eq:Z-annulus} we have used $q_B=e^{2\pi i \tau_B}=\frac{\ell}{L}$ 
(with the modular parameter $\tau_B=-\frac{i}{2\pi} \; \ln \left(\frac{L}{\ell}\right)$).

Similarly, the partition functions on the two disks, regions $A$ and $A\cup B$, are conformally mapped to two infinitely long
cylinders (as the UV cutoff $a \to 0$) with Neumann-Dirichlet boundary conditions. These partition functions are
\begin{equation}
Z_{\rm disk}=2^{-5/12} \pi^{1/4} \frac{\vartheta_4 \left(\tau \right)}{\eta(q^2)},
\label{eq:ZNN}
\end{equation}
where $q=\left(\frac{a}{\ell}\right)^4,\left(\frac{a}{L}\right)^4$ for regions $A$ and $A \cup B$, respectively, 
and $\tau$ is their corresponding modular
parameter; $\vartheta_4(\tau)$ is the elliptic theta-function 
\begin{equation}
\vartheta_4(\tau)=\sum_{n=-\infty}^\infty (-1)^n q^{\frac{n^2}{2}}.
\label{eq:theta4}
\end{equation}

The resulting entanglement entropy for the planar (disk) geometry is found to be
\begin{equation}
S=\frac{1}{2} \; \ln \left[\frac{1}{\pi} \ln \left(\frac{L}{\ell} \right)\right]+  \ln R.
\label{eq:SDisk}
\end{equation}
Hence, for the case of the disk there is also a universal finite piece in the entanglement entropy, 
$\gamma_{QCP}=\ln \sqrt{2kr^2}\equiv \ln R$. 
As in the cases discussed above (the cylinder and the torus), here too $\gamma_{QCP}$ has a
topological origin as it is due to the winding modes of the compactified boson. 
However, unlike the case of the of the cylinder and toroidal geometries, in the case of the disk there is
also a dependence on the aspect ratio $L/\ell$ (the double logarithmic term), as already noted in
Ref.[\onlinecite{Fradkin2006}]. (Note that we included the factor of $1/\pi$ in the double logarithm since it arises from the conformal mapping.)

\subsection{Entanglement Entropy of Quantum Dimer Models and Related Systems}
\label{subsec:micro}

The results on the entanglement entropy of the preceding subsections apply to several ``microscopic'' systems of interest. The
simplest of them is the quantum dimer model on bipartite lattices at the RK point (associated with the RK wave function of the QDM).  
As noted in Ref.[\onlinecite{Ardonne2004}],
the RK point of the QDM maps onto the quantum Lifshitz model for a particular value of the radius $r=1$ and stiffness $k=2$ (in
the notation used here.) This corresponds to a 2D Euclidean boson CFT at the free fermion radius. Of course, this is not an
accident, since in
this case the lattice partition functions can also be computed exactly by pfaffian methods,
\cite{Fisher1961,Samuel1980,Fendley2002} and hence it is a free Dirac fermion system.

Generalized quantum dimer models have
been discussed recently.\cite{Alet2005,Papanikolaou2007b,Castelnovo2005} In these models the wave functions correspond to
 dimer models with weights that depend on the number of dimer pairs on the plaquettes. For a considerable range of values of these
 weights the system remains critical and can also be mapped onto a quantum Lifshitz model, albeit with a different stiffness
 connected with the presence of an exactly marginal operator. Thus, in these models the stiffness varies continuously as a
 function of the microscopic weights. This dependence, discussed in detail in Ref.[\onlinecite{Papanikolaou2007b}], is of course
 non-universal, as it depends on the microscopic structure of the system. Nevertheless, the
 critical exponents have a universal dependence on the stiffness. The same applies to the universal piece of the 
 entanglement entropy $\gamma_{QCP}$, which can be read-off from the results presented in this section.

Similarly, the {\em quantum} eight-vertex model wave function\cite{Ardonne2004} also maps onto a 
free fermion problem for a special choice of weights.\cite{Baxter1982} 
For general values of $k$ the fermions are interacting (see the discussion below) but the effects only 
enter through an exactly marginal operator. The mapping of the quantum 2D eight-vertex model to the quantum Lifshitz model
was shown in detail in Ref.[\onlinecite{Ardonne2004}] where the relation between the stiffness $k$ of the compactified boson 
and the  Baxter weights is given explicitly.  $k$ and the weight $c$ in the Baxter
wave function (along the six vertex line) are related by
\begin{equation}
\frac{\pi}{2k}=\cot^{-1}\sqrt{\frac{4}{c^4}-1}
\label{eq:6v}
\end{equation}
for a boson with compactification radius $r=1$ or, equivalent, an effective radius $R=\sqrt{2kr^2}$.

The results of the preceding subsections on the entanglement entropy for the quantum Lifshitz model apply to the lattice
models almost without change. Once the mapping of the stiffness to the microscopic parameters 
(as in the case of the quantum eight vertex
model) is known, the universal piece,
$\gamma_{QCP}$, can be read-off immediately. The only caveat here is that in lattice models it is impossible to have
closed simply connected regions with smooth boundaries. The resulting paths of the effective coarse grained quantum Lifshitz
model will always have singularities, such as corners, which contribute with a logarithmic dependence to the entanglement entropy
(as discussed in Ref.[\onlinecite{Fradkin2006}]) rendering the finite terms generally non-universal. The cylinder and torus
geometries are exceptional in this sense, and allow for a direct check of these ideas in microscopic models, either through an
exact solution or by means of numerical computations. 

We end this discussion by  giving the results for the universal entanglement entropies $\gamma_{QCP}$  
for the Lifshitz universality class at the free
fermion (or dimer) and Kosterlitz-Thouless transition of the dimer and Baxter (six vertex) wave functions for all three geometries.
(See the summary of Table
\ref{table:entropies}.) At the ``free dimer'' point (the free fermion point of the dimer
models) the stiffness $k=2$ (corresponding to $c^2=\sqrt{2}$ in the Baxter wave function), and the universal term of the
entanglement entropy for a disk geometry is $\gamma_{QCP}^{\rm disk}=\ln \sqrt{2kr^2}=\ln 2$. 
For the cylinder, also at the free dimer point, we also found
$\gamma_{QCP}^{\rm cylinder}= \ln 2$, while for the torus we obtained
$\gamma_{QCP}^{\rm torus}=2\ln 2$.  (Below we will discuss the relation of these results with the {\em topological} entanglement entropy of
the nearby $\mathbb{Z}_2$ topological phase.) 
Away from the free dimer (or fermion) points, the stiffness $k$ changes and so does the
entanglement entropy. Thus, at the Kosterlitz-Thouless transition point of both the dimer and six vertex wave functions (where the
Baxter weight is $c=\sqrt{2}$), the stiffness is $k=1$. 
(At this point the associated $c=1$ CFT has an $SU(2)_1$ Kac-Moody current algebra, and the effective compactification 
radius here is $R=\sqrt{2}$.) The (finite) entanglement entropies now are $\gamma_{QCP}^{\rm torus}=2 \ln \sqrt{2}$,
$\gamma_{QCP}^{\rm cylinder}=0$, and $\gamma_{QCP}^{\rm disk}=\ln \sqrt{2}$. 

\begin{table}[h]
\newcolumntype{Y}{>{\centering\arraybackslash$}m{1cm}<{$}}
\newcolumntype{C}{>{\centering\arraybackslash$}m{2cm}<{$}}
\renewcommand{\arraystretch}{2}
\begin{tabular}{|C||C|C|C|}
\hline
R &{\rm cylinder}&{\rm torus}& {\rm disk}\\
\hline
 2 \ ({\rm RK point}) &  \ln 2 & 2 \ln 2 & \ln 2 \\
\hline
 \sqrt{2} \  ({\rm KT point}) &  \ln \sqrt{2} & 0 & \ln \sqrt{2}\\
\hline
\end{tabular}
\caption{Universal entanglement entropies $\gamma_{QCP}$ of the lattice models in QLM universality class 
in the cylinder, torus, and disk
geometries.  $\gamma_{QCP}$ based on calculations from QLM is quoted at the free fermion point (or RK point) $R=2$, and at the Kosterlitz-Thouless ($SU(2)_1$) point, $R=\sqrt{2}$.}
\label{table:entropies}
\end{table} 

The only caveat  in applying the calculation of $\gamma_{QCP}$ in the QLM to microscopic models is that is impossible to have
closed simply connected regions with smooth boundaries  on a lattice. Hence the resulting paths of the effective coarse grained QLM 
will always have singularities (such as corners) which contribute a finite logarithmic dependence to the entanglement entropy.
\cite{Fradkin2006}
The cylinder and torus
geometries are exceptional in this sense, and allow for a direct check of these ideas in microscopic models, either through an
exact solution or by means of numerical computations.

\section{Generalized conformal QCPs associated with RCFT}
\label{sec:generalizedqcps}

We now generalize the application of Eq.\eqref{eq:SFM} to the computation of the entanglement entropy to more general case of conformal QCPs, specifically those associated whose wave functions have an associated 2D Euclidean  RCFT (a  CFT with a finite number of primary fields.)

\subsection{Entanglement entropy and Boundary Conformal Field theory}

The ground state wave function for a conformal quantum critical point can be expressed as 
Gibbs weight associated with a 2D Euclidean CFT:
\begin{equation}
\Psi_{GS}[\phi]=\frac{1}{\sqrt{Z}} e^{\displaystyle{-S[\phi]/2}}
\end{equation}
as in the case of the QLM discussed in the previous section. Hence there is a one-to-one mapping between 
the norm square of the wave function 
and the partition function of a local 2D Euclidean CFT, and also between the equal-time correlators 
of the operators of the 2D conformal QCP map onto 
and the correlators of primary fields of the 2D Euclidean CFT. 
Furthermore, we will also assume that the
associated Euclidean CFT is {\em unitary} (the $S$-matrix to be defined below
is unitary) and that it is a {\em RCFT}. The restriction to unitary RCFT allows us to exploit well developed technology for this large class of CFTs\cite{Ginsparg:1988nr,yellow}, especially that  of
operator product expansion (OPE) and of {\em modular S-martirx}, in calculation of $\gamma_{QCP}$.

The behavior of RCFTs with specified boundary conditions (especially their partition functions),  
is the subject of boundary conformal field theory, and was discussed extensively by Cardy\cite{Cardy1986a,Cardy1989}. 
We will follow the approach and results of Cardy in this section.
We also need to specify the boundary conditions at the ends of the cylinder, {\it i.e.\/} the {\em boundary states} 
of the boundary CFT.\cite{Cardy1986a} Let us denote these conformal boundary conditions by $(\alpha,\beta)$. 
The associated (conformally invariant) boundary states $\bra{a}$ and $\ket{b}$ can be constructed for each CFT. 
On the other hand, at the common boundary $\Gamma$ between the regions $A$ and $B$, all $n-1$ 
fields must obey {\rm fixed} (`Dirichlet') boundary conditions. As shown by Cardy,\cite{Cardy1986a} this boundary condition is quite generally given
by the boundary state  $\ket{0}$ in the conformal block of the identity ${\bf 1}$.

For simplicity, we will consider here only the geometries of a cylinder (with specific boundary conditions at each end) and a torus. As in
Eq.\eqref{eq:SFM} we will need to compute the free energies of region  
$A$, $B$ and $A \cup B$ with fixed boundary conditions.

The partition function for a RCFT on a cylinder 
of length $L$ and circumference $\ell$, with boundary conditions $a$ and $b$ on the left and right ends respectively, 
$Z_{a/b}$, can be expressed in terms of the characters $\chi_i$ of the RCFT:
\begin{equation}
Z_{a/b}=\sum_j N^j_{ab} \chi_j\left(e^{\displaystyle{-\pi \ell/L}}\right),
\label{eq:Zbc}
\end{equation}
where the integers $N^j_{ab}$ are the fusion constants, the coefficients in the OPE of the RCFT,
\begin{equation}
\Phi_a \times \Phi_b=\sum_j N^j_{ab} \Phi_j.
\label{eq:OPE}
\end{equation}
The Virasoro characters $\chi_j$ are given by the trace over the descendants  $\ket{\Phi_j}$ of the highest weight state, which are obtained by acting on it % $\ket{\Phi_a}$ 
with the Virasoro generators $\hat{L}_{-n}$ ($n>0$):
\begin{equation}
\chi_j(e^{-\pi \ell/L})=e^{{\pi \ell c }/{24 L}} \; 
\textrm{tr}_a\left(e^{-\frac{\pi \ell}{L} \hat{L}_0}\right),
\label{eq:characters}
\end{equation}
where $c$ is the central charge of the CFT, $\hat{L}_0$ is the $n=0$ Virasoro generator. Here the modular parameter is $\tau\equiv i\ell/2L$.
Under a modular transformation $\tau\rightarrow-1/\tau$, which exchanges the Euclidean ``space'' and ``time''  
dimensions of the cylinder ({\it i.e.\/} it flips the cylinder from the ``horizontal'' to the ``vertical'' position), 
the characters transform as
\begin{equation}
\chi_i\left(e^{\displaystyle{-\pi \ell/L}}\right)=S^j_i\; \chi_j\left(e^{\displaystyle{-4\pi L/\ell}}\right),
\label{eq:modular}
\end{equation}
where $S^j_i$ is the {\em modular $S$-matrix} of the RCFT.  The modular $S$-matrix and
the fusion coefficients are related by the Verlinde formula \cite{Verlinde:1988sn}
\begin{equation}
N^j_{ab}=\sum_i \frac{S^i_jS^i_aS^b_i}{S^i_0}.
\label{eq:verlinde}
\end{equation}

The limit of interest here is, once again, $L \gg \ell$. Under a modular transformation, the partition function 
of Eq.\eqref{eq:Zbc} becomes
\begin{equation}
Z_{a/b}=\sum_{i,j} N^i_{ab} \; S^j_i\; \chi_j\left(e^{\displaystyle{-4\pi L/\ell}}\right).
\label{eq:Zbc2}
\end{equation}
In the limit $\frac{\ell}{L} \to 0$, $Z_{a/b}$ is dominated by the 
%lowest energy states (in terms of the 1+1-dimensional 
%Hamiltonian formulation), 
%{\it i.e.\/}
 the descendants of the identity $\bf{1}$ (up to exponentially small corrections). Hence, in this limit,
\begin{equation}
Z_{a/b} \to \sum_i N^i_{ab} \; S_i^0 \; \chi_0\left(e^{-4\pi L/\ell}\right) \to 
e^{{\frac{\pi L c}{6\ell}}} \; \sum_i N^i_{ab} \; S^0_i
\label{eq:lowT}
\end{equation}
and  $\ln Z_{a/b}$ becomes
\begin{equation}
\ln Z_{a/b}=\frac{\pi L c}{6\ell}+\ln g_{ab},
\label{eq:Zbc-g}
\end{equation}
%Here we have 
dropping UV singular (non-universal) terms. The quantity $\ln g_{ab}$ in Eq.\eqref{eq:Zbc-g} is the 
{\em boundary entropy} of a boundary RCFT introduced by Affleck and Ludwig\cite{Affleck1991}, where the ``ground state degeneracy'' 
$g_{ab}$ is given by
\begin{equation}
g_{ab}=\sum_i N^i_{ab} S_i^0.
\label{eq:g}
\end{equation}

Using  Eq.\eqref{eq:SFM}, these standard results  imply that the 
entanglement entropy of the 2D rational conformal QCP
 for a cylindrical geometry(see Fig.\ref{fig:Cylinder}). For boundary conditions $a$ and $b$
  at the two ends associated with regions $A$ and $B$, the entanglement entropy is
\begin{eqnarray}
S&=& -\ln \left(\frac{Z_A^{a0}Z_B^{0b}}{Z_{A\cup B}^{ab}}\right)
\nonumber \\
&=&\mu \ell -\ln \left(\frac{\left(\sum_j N_{a0}^j \; S_j^0\right) \; 
\left( \sum_k N_{0b}^k\;S_k^0\right)}{\sum_l N_{ab}^l\; S_l^0}\right)
\nonumber \\
&=&\mu \ell-\ln \left(\frac{g_{a0} g_{0b}}{g_{ab}}\right),
\label{eq:S-cylinder-rcft}
\end{eqnarray}
where we explicitly used the fact that the state at the common boundary $\Gamma$ should be fixed to be the {\em fixed} BC with boundary state $\ket{0}$.

The result Eq.\eqref{eq:S-cylinder-rcft} provides an explicit way to compute $\gamma_{QCP}$ 
for the entire class of many-body wave functions at QCPs associated with RCFT
in terms of the data of the RCFT:
\begin{equation}
\gamma_{QCP}=-\ln \left(\frac{\left(\sum_j N_{a0}^j \; S_j^0\right) \; 
\left( \sum_k N_{0b}^k\;S_k^0\right)}{\sum_l N_{ab}^l\; S_k^0}\right).
\label{eq:gammaQCP}
\end{equation}
 This is the main result of this section. It shows that $\gamma_{QCP}$ is in general determined by the OPE coefficients $N_{ba}^c$
(which encode the boundary conditions on the partition functions) and by the modular $S$-matrix, $S_i^j$, of the RCFT associated with
the {\em norm squared of the many-body wave function} at the given QCP. 

It is important to note that it is also possible to define a unitary $S$-matrix that governs the transformation properties
of the {\em wave function} itself under a modular transformation. This modular $S$-matrix plays a central role in 2D topological phases and in topological field theories.\cite{Witten1989,Kitaev2006a,Bonderson2006b} However, only for topological theories these are two $S$-matrices are the same and in general they are different or not even defined at all.  We will come back to this issue in the discussion section.

A particularly simple result is obtained for the case of a cylinder with fixed boundary conditions on both ends. In this case, $Z_A$,
$Z_B$ and $Z_{A \cup B}$ are cylinders with fixed boundary conditions, and hence the boundary states for all three cases are in the
conformal block of the identity ${\bf 1}$.  Since in this case the only non-vanishing OPE coefficient is $N_{0 0}^0=1$, the
universal term of the entanglement entropy, $\gamma_{QCP}$, depends only on the element $S_0^0$ of the modular $S$-matrix of the RCFT:
\begin{equation}
\gamma_{QCP}=-\ln S_0^0.
\label{eq:gammaQCP-simple}
\end{equation}

For the case in which the full region$A \cup B$ is a torus,  we can use an analogue of Eq.\eqref{eq:S-cylinder-rcft} 
by writing the partition function $Z_{A \cup B}$  in the denominator of Eq.\eqref{eq:S-cylinder-rcft} 
as a modular invariant.  In the limit of interest $L \gg \ell$,  the denominator $g_{ab}$ of Eq.\eqref{eq:S-cylinder-rcft} 
is replaced by a sum of terms with similar structure corresponding to a sum over boundary conditions (and twists) needed to 
represent the torus (see, for instance, Ref.[\onlinecite{yellow}]). Similarly, Eq.\eqref{eq:S-cylinder-rcft} can also be applied 
to the disk geometry upon a conformal mapping as it was done for the case of the compactified boson in section \ref{subsec:Disk}.

\subsection{Applications}

We will now discuss some examples of interest. 
In applying the results Eq.\eqref{eq:gammaQCP} to specific systems, one should keep in mind that 
that choice of the inner product of the 2D quantum theory can play a subtle role.
As it was pointed out recently by Fendley\cite{Fendley2008}, a scale invariant  wave function 
does not necessarily imply scale invariance of the correlators.  Their actual behavior depends also on the choice of inner product. Here we have assumed that the states labeled by the set of field configurations $\phi(x,y)$ form an orthogonal basis. Hence, the norm of the wave function is a sum over states  with the local weights squared. However what matters is that the {\em matrix elements} (and in particular the norm of the states) be scale-invariant. A number of interesting counterexamples are known.\cite{foot1}
The QLM is a special case where such ``naive'' inner product maintains scale invariance. This is due to the existence of 
exactly marginal operators in the QLM.

Below we discuss four cases where the ground state wave function with the ``naive'' inner product describes QCPs:
(i) a QCP associated with the 2D Ising CFT, (ii) the QCPs associated with compactified boson CFT, (iii) QCPs in quantum loop models\cite{Freedman2004b,Troyer2008}, and 
(iv) quantum net models\cite{Levin2005,Fendley2005,Fidkowski2006,Fendley2008}. (See footnote Ref.[\onlinecite{foot2}].)

\subsubsection{The 2D Ising wave function}
As an example of a system described by an RCFT we consider a 2D quantum spin system whose ground state wave function has for 
amplitudes the Gibbs weights of the 2D classical Ising model. This system is quantum critical if the square of the weights 
(which also have the form of a Gibbs weight for the 3D Ising model) are at  the critical point of the 2D Ising model, 
the Onsager value. 

The critical point of the 2D Ising model is the simplest RCFT. It has central charge $c=1/2$, and three (bulk) primary fields: 
1) the identity ($\bf{1}$, with conformal weight $h=0$), 2) the energy density ($\varepsilon$, with conformal weight $h=1/2$), and 3) 
the spin field ($\sigma$, with conformal weight $1/16$), which obey the operator algebra (OPE)
\begin{eqnarray}
&&\varepsilon \times \varepsilon={\bf 1}\nonumber \\
&&\varepsilon \times \sigma=\sigma \nonumber \\
&&\sigma \times \sigma={\bf 1}+\varepsilon.
\label{eq:Ising-OPE}
\end{eqnarray}
 The critical Ising model has three possible boundary states:\cite{Cardy1986a} 1) the {\em spin up} state $\ket{+}$, 
 2) the {\em spin down} state $\ket{-}$, and 3) the {\em free} state $\ket{f}$. (Either the up or the down state can 
 be regarded as the fixed boundary state.) These three boundary states, $\ket{+}$, $\ket{-}$, and $\ket{f}$ are 
 in the conformal blocks of the identity ${\bf 1}$ (denoted by $\ket{\tilde{0}}$), the energy density $\varepsilon$ (denoted by
 $\ket{\tilde{\frac{1}{2}}}$, and the spin field $\sigma$ (denoted by $\ket{\tilde{\frac{1}{16}}}$), respectively. 
 The boundary states are given by\cite{Cardy1986a}
\begin{eqnarray}
\ket{+}&\equiv&\ket{\tilde 0}=\frac{1}{\sqrt{2}} \ket{0}+\frac{1}{\sqrt{2}}\ket{\varepsilon}+
\frac{1}{\sqrt[4]{2}}\ket{\sigma}\nonumber\\
\ket{-}&\equiv&\ket{\tilde {\frac{1}{2}}}=\frac{1}{\sqrt{2}} \ket{0}+\frac{1}{\sqrt{2}}\ket{\varepsilon}-
\frac{1}{\sqrt[4]{2}}\ket{\sigma}\nonumber\\
\ket{f}&\equiv&\ket{\tilde{\frac{1}{16}}}= \ket{0}-\ket{\varepsilon}.\nonumber\\
&&
\label{eq:boundary-states-Ising}
\end{eqnarray}
The modular $S$-matrix is
\begin{equation}
S=
\left(
\begin{array}{ccc}
\frac{1}{2} & \frac{1}{2} & \frac{1}{\sqrt{2}} \\
\frac{1}{2} & \frac{1}{2} & -\frac{1}{\sqrt{2}} \\
\frac{1}{\sqrt{2}} & -\frac{1}{\sqrt{2}} & 0
\end{array}
\right),
\label{eq:Ising-S}
\end{equation}
where the columns are labeled by the highest weights $0$, $1/2$, and $1/16$, in that order. 

The entanglement entropy for this wave function can now be computed, using the result of Eq.\eqref{eq:S-cylinder-rcft}.
 We will take region $A \cup B$ to be a long cylinder of length $L$ and circumference $\ell$, and regions $A$ and $B$ to be 
 two cylinders of lengths $L_A$ and $L_B$ respectively, with the same circumference $\ell$, and with $L=L_A+L_B$. 

Let us take the boundary conditions at both ends of $A \cup B$ to be free. By a conformal mapping, this maps onto the disk. Back on the cylinder, the free boundary condition is described by the boundary state $\ket{f}$, 
which is in the conformal block of the primary field $\sigma$. 
On the other hand,  at the boundary $\Gamma$ between regions $A$ and $B$, we have the fixed boundary condition, 
the up state $\ket{+}$
%, which is in the conformal block of the identity $\bf{1}$. 
We readily find
\begin{eqnarray}
&&g_{\sigma, 0}=N_{\sigma,0}^\sigma S_\sigma^0=\frac{1}{\sqrt{2}} \nonumber \\
&&g_{0,\sigma}=N_{0,\sigma}^\sigma S_\sigma^0=\frac{1}{\sqrt{2}}\nonumber\\
&&g_{\sigma,\sigma}=N_{\sigma,\sigma}^0 S_0^0+N_{\sigma,\sigma}^\varepsilon S_\varepsilon^0=1.
\end{eqnarray}
The universal term of the entanglement entropy, $\gamma_{QCP}$ now is
\begin{equation}
\gamma_{QCP}=-\ln \frac{g_{a0} g_{0b}}{g_{ab}}=-\ln \frac{\left(S_\sigma^0\right)^2}{S_0^0+S_\varepsilon^0}=\ln 2.
\label{eq:entropy-cylinder-Ising-free}
\end{equation}
On the other hand, we could consider instead the case of fixed boundary conditions at both ends of the cylinder $A \cup B$. This corresponds to the boundary state $\ket{\tilde{0}}$. Since the boundary condition on $\Gamma$ is always {\rm fixed}, 
%and hence also in the conformal block of the identity, 
$\gamma_{QCP}$ is now
\begin{equation}
\gamma_{QCP}=-\ln S_0^0=\ln 2.
\label{eq:entropy-Ising-cylinder-fixed}
\end{equation}
In the case where $A \cup B$ is  torus of large circumference $L$ and small circumference $\ell$ 
(hence with modular parameter
$\tau=i\ell/L$), the regions $A$ and $B$ are cylinders each of length $L_A$ and $L_B$ and circumference $\ell$, with fixed boundary
conditions at both ends.  The partition function for the torus, $Z_{A\cup B}^{\rm torus}$, is\cite{yellow,Ginsparg:1988nr}
\begin{equation}
Z_{A\cup B}^{\rm torus}=\frac{1}{2}\left(\bigg\vert \frac{\vartheta_2(\tau)}{\eta(\tau)}\bigg\vert+\bigg\vert \frac{\vartheta_3(\tau)}{\eta(\tau)} \bigg\vert+ \bigg\vert\frac{\vartheta_4(\tau)}{\eta(\tau)}\bigg\vert\right).
\label{eq:Z-Ising-torus}
\end{equation}
Using the modular invariance of $Z$ on the torus ($\tau \to -1/\tau$), one finds that in the limit $L\gg \ell$, 
$Z_{A \cup B}^{\rm torus}
\to \frac{3}{2}$.  Hence, in the case of the torus, $\gamma_{QCP}$ is
\begin{equation}
\gamma_{QCP}^{\rm torus}=-\ln \frac{\left(S_0^0\right)^2}{\frac{3}{2}}=\ln 6.
\label{eq:gammaQCP-torus}
\end{equation}

\subsubsection{The compactified boson wave function}

We can also use this approach to compute the entanglement entropy for the compactified boson wave function (the quantum Lifshitz
state) discussed in the previous Section. 
However, unlike the explicit computation of the boson determinant presented in the previous section, 
a computation that can be done for any compactification radius $R$, 
the boundary CFT approach we are using in this section only applies for  a rational CFT. 
This restricts the compactification radius to 
be such that $R^2$ is a rational number. (The general case can be regarded as a limit.) 

It is now  straightforward to compute the entanglement entropy using Eq.\eqref{eq:S-cylinder-rcft}. 
For this case we find $\gamma_{QCP}=-\ln S_0^0=\ln R$, consistent with the results of the preceding section.

\subsubsection{Quantum loop models}

Quantum loop models are two-dimensional quantum systems whose Hilbert space is spanned by states labelled by
loop configurations (or coverings) of a two-dimensional lattice. We will denote by $\{\mathcal{L}\}$ the set of these configurations. 
Conventionally, this set of states 
are taken to be a basis of the loop Hilbert space, and hence they are assumed to be linearly independent, complete and orthonormal,
(with respect to the naively defined inner product.) 

Quantum loop models were originally proposed as candidates for time-reversal invariant topological phases.
\cite{Freedman2001,Freedman2004,Freedman2004b} Wave functions in the Hilbert space of (multi) loop configurations have the form
\begin{equation}
\ket{\Psi_{(x,d)}}=\sum_{\mathcal{L}} x^{L[\mathcal{L}]} d^{N[\mathcal{L}]} \ket{\mathcal{L}}.
\label{eq:loop-zd}
\end{equation}
Here $N[\mathcal{L}]$ is the number of loops in state (configuration) $\mathcal{L}$, $L[\mathcal{L}]$ is the length of loop in the
configuration, 
$d$ is the ``loop fugacity'', and $x$ is the weight (fugacity) of a unit length of loop.

The candidate wave functions of a quantum loop model in a putative topological phase 
depends on the loop configuration but not on the length of the loops. The simplest such state is the ``$d$-isotopy'' 
(multi) loop wave function'' \cite{Freedman2001,Freedman2004}
\begin{equation}
\ket{\Psi_d}=\sum_{\mathcal{L}} d^{N[\mathcal{L}]} \ket{\mathcal{L}}
\label{eq:loop-d}
\end{equation}
obtained from $\ket{\Psi_{(x,d)}}$ by setting the fugacity of the unit length of loop $x=1$.
This is a generalization of Kitaev's ``Toric Code'' wave function\cite{Kitaev2003} ($d=1$), {\it i.e.\/} a $\mathbb{Z}_2$ gauge theory
deep in its deconfined phase in $2+1$ dimensions.
Another limit of interest is the ``fully packed'' state
\begin{equation}
\ket{\Psi_{(\infty,d)}}={\lim_{ x \to \infty}} \sum_{\mathcal{L}} x^{L[\mathcal{L}]}d^{N[\mathcal{L}]} \ket{\mathcal{L}}
\label{eq:loop-infty-d}
\end{equation}
obtained by setting $x \to \infty$, which forces the constraint that the loops cover the maximal allowable set of links on the
lattice. 
 
With the naively defined inner product, the norm squared of the $d$-isotopy state $\ket{\Psi_d}$, Eq.\eqref{eq:loop-d}, is
\begin{equation}
Z(d^2)\equiv||\Psi_d||^2=\sum_{\mathcal{L}} d^{2N[\mathcal{L}]},
\label{eq:Zd}
\end{equation}
which is the same as the partition function of a 2D classical loop model on the same lattice, with a weight $d^2$ per loop.  Likewise, the norm
squared of the fully packed loop state $\ket{\Psi_{(\infty,d)}}$ is the partition function $Z(\infty,d^2)$ 
of the classical fully packed loop model, with fugacity $d^2$, on the same lattice.

The partition functions of classical loop models on a 2D lattice have been studied extensively, particularly on the honeycomb lattice
(for a detailed review see Refs.[\onlinecite{Nienhuis1987,Kondev96,Kondev96b}].) In the fully packed limit, the partition function
$Z(\infty,d^2)$ is critical for $d\leq \sqrt{2}$. The universality classes of the fully packed loop models (on the honeycomb lattice)
are rational {\it{unitary}} CFTs only for $d=1$ (the $SU(2)_1$ RCFT) and $d=\sqrt{2}$ (the $SU(3)_1$ RCFT).
For finite $x$, the partition function for the dense loop gas $Z(x,d^2)$ is also critical for $d\leq \sqrt{2}$. 
The universality classes are again rational unitary CFTs only for $d=1$  and $d=\sqrt{2}$. The fixed point for the case $d=1$ is equivalent to the statistics of the proliferated domain walls of the classical 2D Ising model at infinite temperature.\cite{Nienhuis1987} For $d=\sqrt{2}$ the dense and dilute loop gases have the same critical theory, the Kosterlitz-Thouless critical point, and hence also the $SU(2)_1$
RCFT. 
 
We can now use the result in Eqs.\eqref{eq:gammaQCP} and \eqref{eq:gammaQCP-simple} to compute the universal term of 
the entanglement entropy for the loop wave functions with $d=1,\sqrt{2}$, on a cylinder with fixed boundary conditions (for the loops). The modular $S$-matrices are
known,\cite{Ginsparg:1988nr,yellow,Dong2008} and the needed $S_0^0$ matrix elements are 
$S_0^0=\frac{1}{\sqrt{2}},\frac{1}{\sqrt{3}}$, for
$SU(2)_1$ and $SU(3)_1$, respectively. The universal term $\gamma_{QCP}$ of the entanglement entropy for each case is
$\gamma_{QCP}=\ln \sqrt{2}, \ln \sqrt{3}, -\ln 2$ for the fully packed state at $d=1$ (and also for the loop gas at $d=\sqrt{2}$), the fully
packed loop state at $d=\sqrt{2}$, and the dense loop gas at $d=1$ (corresponding to the Kitaev state), respectively.
Here we have used a recent result on the behavior of of the dense loop model by Cardy\cite{Cardy2006} who showed (among many other things) that for $d=1$ the partition function of the dense loop model on the cylinder $Z=2$. We will see in the discussion section that this {\em negative} value, $\gamma=-\ln 2$, coincides with the direct computation of the {\em topological} entanglement entropy in the Kitaev wave function.\cite{Hamma2005,Levin2006,Kitaev2006a}

\subsubsection{Quantum net models}

Finally, we will briefly discuss the more interesting, but less understood problem of the wave functions for {\em quantum net models}\cite{Levin2005,Fendley2005,Fidkowski2006,Fendley2008}.  These states were proposed as candidates for a time-reversal invariant non-Abelian topological phase. The Hilbert space of quantum net models is spanned by the coverings of a lattice by configurations of nets, {\it i.e.\/} branching loops (with trivalent vertices).  An interesting example is the chromatic polynomial state.\cite{Fendley2005} In this state, the nets are regarded as a configuration of  domain walls of a $Q$-state Potts model. The weight of a given state $\ket{\mathcal{L}}$ is the chromatic polynomial $\chi_Q[\mathcal{L}]$ of the configuration. The chromatic polynomial counts the number of ways of coloring regions of the lattice separated by domain walls of a $Q$-state 2D Potts model. They were first introduced in the computation of the low temperature expansion for the 2D Potts models (see, for instance, Ref.[\onlinecite{Baxter1982}].) For non-integer $Q$, the chromatic polynomial can be computed by an iterative procedure.\cite{Fendley2005} The 2D Potts model is known to have a critical point for $Q\leq 4$.

Following Ref.[\onlinecite{Fendley2005}], we consider the norm of the chromatic polynomial state with $Q\leq 4$. In order to compute the norm, we have to square the weight, resulting in a partition function involving the sum of the {\em square} of the chromatic polynomial. It is then natural to ask for a value of $Q$ such that $\chi_Q^2[\mathcal{L}] \propto \chi_{Q_{{\rm eff}}}[\mathcal{L}]$, for some $Q_{\rm eff}$. Then the nets will be critical provided $Q_{{\rm eff}} \leq 4$. It turns out\cite{Fendley2005} that, up to  a suitably chosen fugacity for trivalent vertices\cite{Fidkowski2006}, this property holds only for $\sqrt{Q}=\frac{1+\sqrt{5}}{2}$, the {\em Golden Ratio}, with $Q_{\rm eff}=2+\frac{1+\sqrt{5}}{2}<4$. Thus, for this state the nets are critical.

This case is interesting for several reasons. One is that strong arguments\cite{Fendley2005} suggest that it is possible to define for this wave function an excitation (a defect)  which is denoted by $\tau$, a Fibonacci anyon (not to be confused with the modular parameter!) with the fusion rule, $\tau \times \tau={\bf 1} + \tau$. Fibonacci anyons are of prime interest in the topological approach to quantum computation.\cite{Freedman2002} However, for this approach to work it is necessary that this state should describe a topological state, which requires that its local excitations (not the nets) be gapped.  Fendley\cite{Fendley2008} has recently given strong arguments that imply that this state, with the naive inner product we use here, is not topological but a quantum critical state. 

Another feature that makes this state interesting is that the correlations encoded in the norm of the state for $\sqrt{Q}=\frac{1+\sqrt{5}}{2}$ are described by a RCFT, the minimal model of the Friedan-Qiu-Shenker\cite{Friedan1984} series of unitary RCFTs at level $m=9$, with central charge $c=\frac{14}{15}$. This minimal model has a large number of primaries (36) and has not been studied in detail.  Nevertheless, its modular $S$-matrix is known (as it is for the entire series\cite{Ginsparg:1988nr}). Although to the best of our knowledge the boundary CFT of this minimal model has not been investigated, we conjecture that the boundary state corresponding to the fixed boundary condition is the analog of the state $\ket{\tilde 0}$ in the 2D critical Ising model (the $m=3$ member of the same series.), {\it i.e.\/}  the state in the conformal block of the identity.\cite{Cardy1989} Thus, if we consider this state on a cylinder with fixed boundary conditions, the entanglement entropy for observing only half of the system, has a universal term $\gamma_{QCP}$ of the form given in Eq.\eqref{eq:gammaQCP-simple},  and hence is given in terms of the $S_0^0$ element of the modular $S$-matrix of this RCFT:\cite{Ginsparg:1988nr}
\begin{equation}
\gamma_{QCP}=-\ln S_0^0=-\ln \left(\frac{\sin (\frac{\pi}{9})}{15+3\sqrt{5}}\right).
\label{eq:gammaQCP-fibonacci}
\end{equation}

\section{Conclusions and Discussion}
\label{sec:conclusions}

We have shown that at 2D conformal QCPs (with dynamical exponent $z=2$), the entanglement entropy  for a region with a smooth boundary quite generally has universal finite contributions  which we denoted by $\gamma_{QCP}$:
\begin{equation}
S_{QCP}=\mu \ell+\gamma_{QCP}.
\nonumber
\end{equation}
We studied the universal nature of $\gamma_{QCP}$ with two complementary approaches  for large classes of 2D conformal QCPs: 
First for the QLM universality class, we calculated  $\gamma_{QCP}$  explicitly in terms of 
the partition functions (that of compactified boson) associated with the norm squared of the wave function. 
 Later we used known results from boundary CFT to show  that $\gamma_{QCP}$ is determined by the detailed structure of the associated RCFT encoded in the modular $S$-matrix and the OPE fusion coefficients for the primary fields.
  We also applied this general results to compute $\gamma_{QCP}$ in several systems of interest: the quantum Lifshitz model, the generalized quantum dimer and quantum eight-vertex models, and quantum loop and net models.

However, we showed (c.f. Eq.\eqref{eq:gammaQCP-simple}) that for a general conformal quantum critical point, whose ground state wave function is given by the Gibbs weights of a Euclidean rational unitary CFT, the universal term $\gamma_{QCP}$ is determined by the modular $S$-matrix  associated with the norm squared of the wave function. Thus, the modular $S$-matrix of the topological phase and that of  the wave functions of 2D conformal quantum critical points have a conceptually different origin.  
%In particular, in all the cases we checked here, $\gamma_{QCP}$ and $\gamma_{\rm topo}$ contribute with opposite signs to their 
%respective entanglement entropies.

We note that while our result for the entanglement entropy has the {\em same form} as the entanglement entropy for a {\em topological phase}, \cite{Kitaev2006a,Levin2006} the finite universal terms $\gamma_{QCP}$ and $\gamma_{\rm topo}$  have a different origin and structure. In the case of a topological phase, $\gamma_{\rm topo}$ is in general determined by the modular $S$-matrix of the 
topological field theory of the topological phase, and it is given in terms of topological invariants of the effective topological field theory that describes this phase.\cite{Kitaev2006a,Levin2006,Dong2008} 
This modular $S$-matrix governs the transformation properties of the ground state within the degenerate ground state Hilbert space of the topological phase
under modular transformations on a torus: $\tau \to -1/\tau$, where $\tau$ is the modular parameter of the torus\cite{Witten1989}. 
On the other hand, for 2D conformal QCPs whose ground state wave function is given by the Gibbs weights of a Euclidean rational unitary CFT, 
 the universal term $\gamma_{QCP}$ is determined by the modular $S$-matrix  associated with the norm squared of the wave function and the $S$-matrix connects between different boundary conditions. Hence the roles of the modular $S$-matrix in the computation of the universal $\mathcal{O}(1)$ terms to the entanglement entropy have 
 conceptually different origin. Moreover, $\gamma_{QCP}$ and $\gamma_{topo}$ enter with opposite signs in their contributions to their respective entanglement entropies.
 In fact, in all the cases we looked at we found that $\gamma_{QCP}>0$, except for the Kitaev state which is topological, and we recovered the known result. (It is unclear to us how general this difference actually is and, more importantly, if it has a deeper meaning.) In any case, the fact that the entanglement entropy has the universal form of Eq.\eqref{eq:Stopo} has led to the widespread assumption that this scaling is a signature of a topological phase. However we have shown here that this is not necessarily the case as this scaling is also obeyed at conformal quantum critical points in 2D.

It is also interesting to note the striking similarity of the structure of Eq.\eqref{eq:gammaQCP} (with its dependence on the $S$-matrix and the fusion rules) with the results of Fendley, Fisher and Nayak\cite{Fendley2007c} for the change in the entanglement entropy of a 2D topological fluid, a fractional quantum Hall state, by the action of a point contact. Recently, Refs.[\onlinecite{Caraglio2008,Furukawa2008}] found finite universal terms in the entanglement entropy for $1+1$ dimensional CFTs with a similar structure to what we found here in 2D  conformal QCPs. Calculations of quantum fidelity in 1D also find a similar structure.\cite{Abasto2008,Venuti2008} Recent work by Li and Haldane\cite{Haldane2008} also raises the interesting possibility of computing the entanglement spectrum for a theory with a wave function described by a known CFT, but this is beyond the scope of this paper.  

Finally, given the close connection between the universal piece of the entanglement entropy $\gamma_{QCP}$ and the Affleck-Ludwig entropy of the associated 2D classical partition functions it is interesting to inquire if $\gamma_{QCP}$ may flow under some perturbation. Clearly this cannot happen under the action of a {\em boundary perturbation} (as in the Affleck-Ludwig case) as that would require one to make a physical change of the wave function on the boundary $\Gamma$, rather than a measurement.  However,  it is interesting to consider instead how the entanglement entropy (and in particular the finite term $\gamma_{QCP}$) would 
evolve as one perturbed the (bulk) system either by a finite non-zero temperature into the quantum critical regime,
 or by a relevant operator that drives the system into a nearby topologically ordered phase that can be accessed by local perturbations
 \cite{Moessner2001,Fendley2002,Ardonne2004,Fendley2005,Fendley2008} and to investigate possible connections with RCFT.\cite{Dorey:1999cj, Dorey:2004xk, Green:2007wr}

\begin{acknowledgments}
We thank John Cardy, Paul Fendley, Greg Moore, and Joel Moore for their comments and suggestions. BH and MM thank the 
Les Houches Summer 
School for its hospitality. The work of EF and BH was supported by the
National Science Foundation Grant No. DMR 0758462 and DMR 0442537 at the University of Illinois.  MM was supported by the Stanford Institute for Theoretical Physics, the NSF under grant PHY-0244728, the DOE under contract DE-AC03-76SF00515, and the ARCS Foundation. EAK was supported by the Stanford Institute for Theoretical Physics during a part of this work.
\end{acknowledgments}

%bibliography{biblio2.bib}

\end{document}